\documentclass[journal]{IEEEtran}

\usepackage{graphicx}
\usepackage{cite}
\usepackage{url}
\usepackage{booktabs}
\usepackage{xcolor}
\usepackage{amsmath}
\usepackage{amssymb}
\usepackage{balance}
\usepackage[normalem]{ulem} 
\begin{document}
	
	
	\title{Hollow-Core Fiber in Direct-Detection Optical Networks: Technology Readiness, Deployment Drivers, and Adoption Outlook}
	\author{Md Ghulam Saber and Zhiping Jiang%
		\thanks{M. G. Saber and Z. Jiang are with the Ottawa Research Center, Huawei Technologies Canada, 303 Terry Fox Drive, Kanata, ON, K2K 3J1,  Canada. e-mail:md.ghulam.saber@huawei.com.}%
	}

	\markboth{Journal of \LaTeX\ Class Files,~Vol.~xx, No.~x, xxx~2026}%
	{Shell \MakeLowercase{\textit{et al.}}: Bare Demo of IEEEtran.cls for IEEE Journals}
	\maketitle
	
	\begin{abstract}
		This paper presents a comprehensive analysis of hollow-core fiber (HCF) for
		intensity-modulation and direct-detection (IMDD) optical networks, covering
		fiber-level physics, system-level performance, and deployment economics.
		We quantify the three principal advantages of anti-resonant HCF over standard
		single-mode fiber (SMF) for IMDD: (i)~chromatic dispersion of 2--4~ps/(nm$\cdot$km)
		versus 17~ps/(nm$\cdot$km), which shifts the first dispersion-induced power-fading
		null from $\sim$10~GHz to 20--28~GHz at 40~km, extending the dispersion-limited
		reach by 4--8$\times$; (ii)~a nonlinear coefficient approximately 1,000$\times$
		lower than silica, permitting launch powers of +10 to +20~dBm and yielding
		7--17~dB of additional link budget; and (iii)~a group index near unity
		($n_g \approx 1.003$), reducing propagation latency by 31\%.
		We further analyze inter-modal interference (IMI) as the dominant impairment for
		HCF-based IMDD. We show that differential modal attenuation (DMA) exceeding
		12~dB/km suppresses IMI-induced crosstalk below the $-$30~dB multipath interference
		threshold required for PAM4.
		The reduced dispersion also lowers the required feed-forward equalizer (FFE)
		tap count by 3--6$\times$, directly decreasing noise enhancement penalty and
		DSP complexity.
		A deployment cost model across five application scenarios---intra-data center,
		campus DCI, metro DCI, 5G fronthaul, and PON---reveals that fiber cable
		constitutes only 5--10\% of outside-plant deployment cost, and that coherent
		transceiver avoidance savings of \$1,000--2,000 per transceiver can offset the
		current HCF premium at metro distances. We provide a technology adoption
		roadmap indicating that HCF is economically justified now for intra-DC and
		campus DCI, with metro DCI following in 2027--2030 as manufacturing costs
		continue to decline.
	\end{abstract}
	
	\begin{IEEEkeywords}
		Hollow-core fiber, intensity modulation and direct detection, anti-resonant
		fiber, inter-modal interference, multipath interference, data center interconnect, deployment economics,
		chromatic dispersion, feed-forward equalization.
	\end{IEEEkeywords}
	
	\section{Introduction}
	
	\IEEEPARstart{I}{ntensity} modulation and direct detection (IMDD) is the
	dominant technology for short-reach optical networks, spanning intra- and
	inter-data center interconnects (DCI), 5G fronthaul, and passive optical
	networks (PON). Industry standards from IEEE~802.3 (400GbE, 800GbE) and the
	Optical Internetworking Forum (OIF) specify IMDD with PAM4 modulation for
	reaches up to 10~km, while emerging 1.6TbE specifications target 2~km. By
	eliminating the local oscillator, polarization diversity receiver, and
	associated digital signal processing for carrier recovery and chromatic
	dispersion (CD) compensation, IMDD transceivers achieve significantly lower
	cost, power consumption, and footprint than their coherent counterparts---a
	typical 400G IMDD module consumes 7--12~W, whereas modern 400G coherent pluggables (400ZR) require 15--18~W.
	
	This simplicity, however, comes at a fundamental cost. Because the square-law
	photodetector recovers only the intensity envelope of the optical field,
	discarding all phase information, IMDD systems cannot compensate for chromatic
	dispersion electronically. The resulting power-fading transfer function
	exhibits periodic spectral nulls whose frequencies depend on the accumulated
	dispersion, and these nulls are irrecoverable without phase information. As symbol rates increase to support 100G, 400G, and 800G PAM4 signaling at 53--112~GBaud, this dispersion-induced fading---not fiber attenuation---becomes the binding reach constraint, rapidly destroying the channel at metro distances. For next-generation 400G and 800G transceivers targeting DCI distances of
	10--40~km, conventional fiber simply cannot support IMDD without either
	costly coherent detection or dispersion-compensating modules.
	
	Hollow-core fiber (HCF) addresses this limitation at the physical layer. This paper focuses on antiresonant hollow-core fiber (AR-HCF). Throughout the paper, HCF and AR-HCF are used interchangeably to refer to AR-HCF.
		Photonic bandgap hollow-core fiber (HC-PBGF), a related but distinct design
		with narrower bandwidth and higher loss, is discussed
		only as a historical reference.
	In AR-HCF, light propagates through an air core rather than
	silica, reducing CD to only 2--4~ps/(nm$\cdot$km)---roughly one-seventh that
	of SMF---while simultaneously lowering the nonlinear coefficient by
	approximately three orders of magnitude and reducing one-way latency by 31\%
	through a near-unity group index. These are not incremental improvements;
	they fundamentally alter the dispersion-limited reach of IMDD systems,
	pushing the onset of fading from single-digit to tens of kilometers.


	Equally important is that HCF has matured beyond the laboratory. The loss
	record now stands at 0.040~dB/km, achieved by YOFC's gap-tube-assisted
	support-tube design with intermodal interference below
	$-$63.6~dB/km.
	Production deployments exceed 1,200~km~\cite{microsoft2024}, with Microsoft
	targeting 15,000~km by late 2026 through partnerships with Corning and Heraeus.
	Commercially available fusion splicers achieve mean splice losses of
	0.043~dB with 100\% success rates~\cite{feng_ofc2026}, and multiple
	manufacturers are now scaling fiber production. Of particular relevance to IMDD
	applications, is the demonstration of a simplified bidirectional PON over
	22~km of AR-HCF supporting 200~Gb/s downstream and 50~Gb/s upstream, with
	power budgets surpassing ITU-T N2 class requirements.

	\begin{figure*}[!t]
		\centering
		\includegraphics[width=\textwidth]{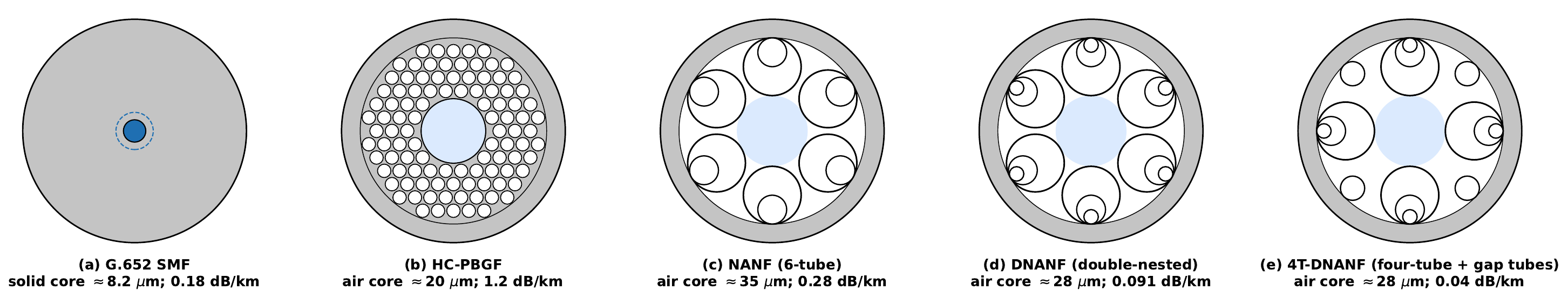}
		\caption{Structural evolution of hollow-core fiber, with standard solid-core
			SMF shown for reference. (a)~G.652 SMF guides light in a small Ge-doped
			solid core ($\sim$8.2~$\mu$m; the dashed ring marks the $\sim$10~$\mu$m
			mode-field diameter). (b)~Photonic-bandgap HCF (HC-PBGF): a periodic
			air-hole lattice confines light to a central air core ($\sim$20~$\mu$m),
			with a minimum reported loss of $\sim$1.2~dB/km. (c)~Nested antiresonant
			nodeless fiber (NANF): six cladding tubes, each with a single nested tube
			tangent to its parent on the side away from the core. (d)~Double-nested
			antiresonant nodeless fiber (DNANF): a second nested tube per cladding
			tube, reaching 0.091~dB/km. (e)~Four-tube DNANF (GTA-DNANF),
			reaching 0.04~dB/km. The air core grows to
			$\sim$20--35~$\mu$m for HCF (mode-field diameter $\sim$20--24~$\mu$m versus
			$\sim$10~$\mu$m for SMF): larger cores suppress confinement loss but admit
			higher-order modes and require mode-field adaptation at HCF--SMF transitions.}
		\label{fig:comparison}
	\end{figure*}
	
	Coherent detection resolves the CD limitation through full-field recovery but
	at substantially higher cost (\$1,500--2,500 versus \$200--400 per
	IMDD transceiver), power, and complexity. For the vast
	majority of short-reach links where IMDD is deployed, this overhead is
	unjustified. HCF offers an alternative: preserving IMDD's cost and simplicity
	while removing its fundamental dispersion constraint.
	
This paper presents a systematic analysis of HCF-enabled IMDD systems. We
characterize the physical advantages of air-core propagation
(Sections~II--III), analyze multipath interference as the primary HCF-specific
impairment (Section~IV), quantify equalization requirements and benefits
(Section~V), and develop a deployment cost model spanning five application
scenarios (Section~VI). Section~\ref{sec:challenges} also provides a dedicated treatment of challenges and uncertainties—connector maturity, field operability, modal-purity and CO\textsubscript{2}-absorption constraints, monitoring limitations, and economic sensitivity—to complement the technical discussion. Section~\ref{conclusion} concludes with an adoption roadmap.

The contribution of this work is an integrative, system-level synthesis rather than a new experimentally validated architecture. Prior literature has treated HCF progress either at the fiber-device level (loss, modal purity, splicing) or as point demonstrations of specific transmission experiments. We connect these threads specifically to IMDD networking constraints—dispersion-limited reach, equalizer complexity, launch-power budget, and latency—and embed them in a deployment-economics framework across five application scenarios. To aid the reader in distinguishing established results from projections, we explicitly flag the epistemic status of each quantitative claim throughout the paper: (i) \textit{reported experimental results} from published papers; (ii) \textit{analytical scaling arguments} derived from first-principles physics, which are exact within their stated assumptions but require experimental validation in specific system contexts; and (iii) \textit{forward-looking projections}, which are scenario-based extrapolations from current manufacturer announcements and experience-curve arguments and should be read as illustrative rather than predictive.
	
	\section{Hollow-Core Fiber Technology Overview}
	
	Conventional optical fibers guide light through a solid silica core via total
	internal reflection. Anti-resonant hollow-core fibers confine light in an air
	core surrounded by a microstructured glass cladding composed of thin-walled
	capillary tubes. The capillary walls act
	as Fabry--P\'{e}rot resonators: the resonance wavelengths are given by
	$\lambda_m = 2t\sqrt{n^2 - 1}/m$, where $t$ is the capillary wall thickness,
	$n \approx 1.45$ is the refractive index of silica, and $m$ is the resonance order. At wavelengths between
	resonances---the anti-resonance condition---reflectivity at the air-glass
	interface approaches unity, and light is strongly confined to the hollow core.
	For C-band operation at 1550~nm, wall thickness of 480--550~nm positions the
	first-order resonance below the operating wavelength, producing broad low-loss
	transmission windows spanning 100--200~nm that cover the full C- and L-bands.
	The loss mechanisms in AR-HCF differ fundamentally from SMF: rather than
	Rayleigh scattering from glass density fluctuations, the dominant contributions
	are confinement loss (leakage through the cladding structure) and surface
	scattering from capillary wall roughness. Confinement loss
	scales as $({\lambda}/{R_\text{core}})^4$, establishing a design tension:
	larger core radii reduce confinement
	loss but support additional higher-order modes, increasing multipath interference (MPI) susceptibility.
	
	Three generations of AR-HCF designs have produced rapid performance
	improvements over the past six years (Fig.~\ref{fig:comparison}). The nested
	anti-resonant nodeless fiber (NANF), proposed by Poletti in 2014,
	employs six capillary tubes each containing a smaller nested element, providing
	multiple layers of anti-resonant confinement. Later on, loss of
	0.65~dB/km was achieved across the C+L bands with this geometry in 2019. The double-nested
	anti-resonant nodeless fiber (DNANF) introduces a second nesting level inside
	each capillary, further suppressing confinement loss and achieving 0.091~dB/km~\cite{slavik_ofc2026}---the lowest reported
	average loss in any optical fiber and the first HCF to surpass the Rayleigh
	scattering floor ($\sim$0.14~dB/km at 1550~nm) that represents the fundamental
	loss limit of solid-core silica fibers. The Rayleigh limit arises from density
	fluctuations frozen into the glass during fiber drawing---a constraint that
	cannot be circumvented in solid-core geometries. HCF avoids this barrier
	entirely by confining light in air. In a parallel advance, Gao et al.\
	demonstrated the four-tube DNANF (4T-DNANF)~\cite{gao2025}, which achieves
	greater control over the trade-off between loss and modal purity.
	Two fiber variants were reported: one with fundamental-mode (FM) loss of
	0.1~dB/km and DMA of 430~dB/km for the LP$_{11}$ mode, and another with FM
	loss of 0.13~dB/km and DMA exceeding 6,500~dB/km---the latter yielding a
	higher-order mode extinction ratio (HOMER) of 50,000, the highest reported to
	date. This ultrahigh DMA is critical for IMI suppression as analyzed in
	Section~IV.
	
	The current minimum reported loss of 0.040~dB/km and the lowest average loss of
	0.091~dB/km~\cite{slavik_ofc2026} both surpass commercial SMF performance.
	Production HCF now offers 0.10--0.15~dB/km, and YOFC has demonstrated a
	266~km spliced link at 0.10~dB/km (including splices) with IMI of
	$-$68.8~dB/km. The trajectory indicates that HCF
	attenuation is no longer a barrier to deployment. An
	additional practical consideration is bend sensitivity: AR-HCF requires
	minimum bend radii of 15--30~mm for acceptable loss, compatible with standard
	data center cable routing but more constrained than SMF. Current DNANF designs
	achieve acceptable bend loss at 15~mm of bend radius, suitable for
	structured cabling and patch panels.

		\begin{figure*}[!h]
		\centering
		\includegraphics[width=\textwidth]{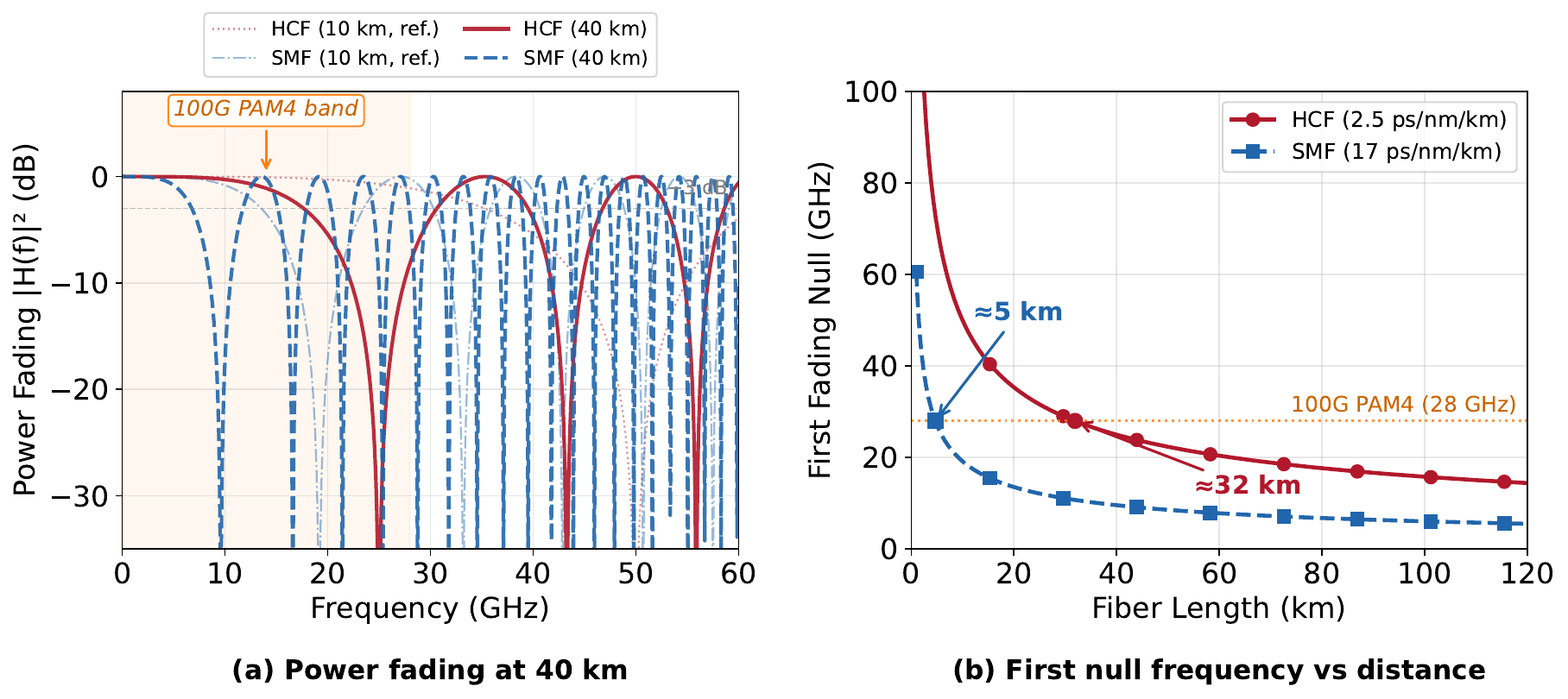}
		\caption{Dispersion-induced power fading in IMDD. (a)~Power fading transfer
				function at 40~km for HCF and SMF, with 10~km shown for reference as faint traces; the 10~km SMF trace, in particular, has its first fading null near $\sim$19~GHz and is shown with a lighter line weight so as not to obscure the 40~km comparison.
				SMF fading nulls fall within the 100G PAM4 signal band at metro distances,
				while HCF remains largely transparent. (b)~First fading-null frequency
				versus fiber length. HCF pushes the first null beyond the 28~GHz PAM4
				bandwidth out to $\sim$32~km, compared with $\sim$5~km for SMF in C-band.}
		\label{fig:dispersion}
	\end{figure*}
	The HCF ecosystem has matured correspondingly. Manufacturers like
	Microsoft/Lumenisity, Corning, Heraeus Covantics, OFS/Furukawa, YOFC, and Relativity Networks---are scaling production capacity. Field-deployable
	automated splicing using standard commercial splicers (e.g., Fujikura FSM-100P)
	now achieves 0.043~dB mean loss for HCF-to-HCF splices with 100\%
	success rates across 30~trials~\cite{feng_ofc2026}, while HCF-to-SMF
	interconnection losses as low as 0.08~dB (0.12~dB for double-nested fiber)
	have been demonstrated via perfect mode-field-size adaptation. HCF-to-SMF
	interconnection with
	back-reflection below $-$60~dB~\cite{slavik_ofc2026} has been demonstrated as well. Microsoft has deployed over 1,200~km of HCF in
	production networks with 15,000~km planned~\mbox{\cite{microsoft2024}}; Amazon Web Services (AWS) has publicly disclosed HCF deployment at a small number of availability-zone interconnect sites as of early 2026~\mbox{\cite{aws_dck_2026}}, primarily for latency-constrained long-distance interconnects between data centers within a single availability zone; and YOFC and China Mobile have deployed commercial HCF links in China. ITU-T Study Group 15 is actively developing technical reports and
	recommendations related to hollow-core fiber, with work currently
	ongoing, while the China Communications Standards Association (CCSA)
	and other regional bodies have initiated activities related to HCF
	test methods. IEC SC86A is expected to provide a framework for fiber
	and cable specifications as the technology matures. Connector
	technology has demonstrated 0.1--0.3~dB loss in laboratory settings, though
	commodity connectors are not yet available. A key interconnection challenge is
	the mode field diameter (MFD) mismatch between HCF (typically 18--25~$\mu$m, approximately 70\% of the core diameter) and SMF ($\sim$10~$\mu$m), which necessitates mode-field adapters or intermediate
	graded-index fiber segments at each transition point. 

		Practical deployability depends strongly on the outer geometry of the cabled fiber, as duct and microduct sizing is a first-order cost driver in both greenfield and brownfield scenarios. Recent AR-HCF/NANF designs increasingly target a standard 125~$\mu$m cladding diameter, with telecom-oriented cores typically in the 20--35~$\mu$m range, enabling compatibility with the footprint of G.652 SMF and the use of conventional 250~$\mu$m coatings, 900~$\mu$m buffers, and loose-tube constructions.
		
		This dimensional alignment allows HCF to be deployed in existing microduct infrastructure (e.g., 7/3.5~mm and larger), subject to installation constraints such as duct occupancy, routing geometry, and allowable pulling forces. In brownfield scenarios, mechanical limits---including bend radius (typically 15--30~mm) and sensitivity to microbending and lateral compression---are often more restrictive than outer diameter. For greenfield builds, the near-identical geometry enables reuse of standard duct sizing, handling, and enclosure practices, reducing incremental civil-works overhead.

		A critical HCF-specific impairment is the absorption fingerprint of ambient gases---primarily CO\textsubscript{2} and water vapor---which can ingress if the fiber lacks hermetic sealing. Measurements demonstrate that CO\textsubscript{2} introduces severe, narrow absorption peaks centered specifically in the L-band which is not typically used for IMDD applications. While this transmission penalty is negligible for short intra-DC spans ($<$10~km), it scales linearly with length, becoming a major constraint for extended-reach metro links. This impairment can be mitigated by purging the core with an inert gas prior to hermetic end-capping during cable production although the process is time consuming. Alternatively, for unsealed deployments, digital signal processing (DSP) techniques like waveform-level pre-compensation, spectral avoidance etc. have been proposed to partially offset the penalty. Consequently, ambient gas absorption remains primarily an extended-reach L-band concern rather than a fundamental barrier for short-haul adoption.

		AR-HCF is not perfectly polarization invariant: twisting and bending perturb the capillary geometry and induce local birefringence and dichroism, with twist producing mainly circular birefringence (two orders of magnitude smaller than in SMF) and bend producing mostly linear birefringence and polarization-dependent loss. For IMDD links, this is usually a secondary concern because direct detection is insensitive to SOP, but it can matter for polarization-sensitive systems such as coherent links and sensing. 

AR-HCF exhibits $\sim$15--30~dB weaker backscatter than SMF because the guided field has little overlap with the glass, and the residual backscatter originates from surface scattering at the microstructure walls and from gas in the core rather than from bulk silica Rayleigh scattering. Backscatter is therefore also non-uniform along the fiber, varying with local core geometry and gas pressure. Conventional single-ended OTDR traces are consequently noisy and can misreport loss, and HCF-to-HCF splices introduce extended ``gas-filling events'' that further distort traces. Commercial HCF-optimized OTDRs with high dynamic range (e.g., $\sim$46~dB) and bi-directional post-processing---which averages out the geometry- and pressure-induced backscatter variation---became available from EXFO and VIAVI in 2025--2026 and are sufficient for fault location, splice characterization, and link certification over distances up to $\sim$150~km. Distributed acoustic/temperature sensing on HCF, and field-deployed multi-vendor operational tooling beyond fault location, remain active areas of work.
	
	\begin{table*}[!h]
		\centering
		\caption{Selected performance benchmarks for anti-resonant hollow-core fiber (as of March, 2026). YOFC: Yangtze Optical Fiber and Cable Joint Stock Limited Company, OFC: Optical Fiber Communication Conference, MWC: Mobile World Congress, GDUT: Guangdong University of Technology, GTA-ST-HCF: Gap Tube Assisted Support Tube HCF }
		\label{tab:advancements}
		\renewcommand{\arraystretch}{1.3}
		\begin{tabular}{@{}p{4.5cm} p{5cm} p{6.5cm}@{}}
			\toprule
			\textbf{Metric} & \textbf{Best Value} & \textbf{Source} \\
			\midrule
			Minimum loss & 0.040~dB/km & YOFC GTA-ST-HCF (OFC 2026) \\
			Longest draw length & 91.2~km & YOFC HollowBand\textsuperscript{\textregistered} HCF (MWC Barcelona 2026) \\
			Wideband low loss & 0.091~dB/km at 1{,}550~nm, $<$0.2~dB/km over 66~THz & Southampton/Microsoft DNANF (\emph{Nature Photonics}, Sep.\ 2025) \\
			Lowest IMI & $-73.2$~dB/km & YOFC GTA-ST-HCF (OFC 2026) \\
			Best splice loss (HCF-HCF) & $<$0.05~dB, 97~seconds & GDUT/China Telecom (OFC 2026) \\
			Best splice loss (HCF-SMF) & 0.08~dB (0.12~dB for 5-ring DNANF); return loss $<-$60~dB & Zhong et al., JLT 2024; Southampton (OFC 2026) \\
			Max BiDi capacity & 2$\times$64$\times$400G over 107.5~km & Microsoft (OFC 2026) \\
			Deployed live traffic & 1{,}200+~km & Microsoft Azure \\
			Planned deployment & $>$12{,}000~km deployment planned & Microsoft Azure \\
			\bottomrule
		\end{tabular}
	\end{table*}
	
Table~\ref{tab:advancements} consolidates the key fiber-, component-, and deployment-level milestones. These entries summarize the state-of-the-art as of early 2026.

	\section{HCF Advantages for IMDD Systems}
	
	The propagation characteristics of HCF address the three principal limitations
	of IMDD systems: dispersion-induced power fading, nonlinearity-constrained
	launch power, and propagation latency.

	\subsection{Mitigation of Dispersion-Induced Power Fading}
	
	In IMDD systems, CD induces a frequency-dependent power fading effect. As the
	modulated optical signal propagates through dispersive fiber, different
	spectral components accumulate differential group delay, producing a periodic
	transfer function at the square-law detector with nulls at frequencies where
	complete signal cancellation occurs. These nulls are irrecoverable: unlike
	coherent detection, where the full electric field is digitized and CD can be
	compensated digitally, the IMDD receiver has no access to the phase information
	needed for post-detection correction.
	
	Figure~\ref{fig:dispersion} illustrates these effects quantitatively: Fig.~\ref{fig:dispersion}(a) shows the 
	power-fading transfer function for HCF and SMF at 10~km and 40~km, and Fig.~\ref{fig:dispersion}(b) plots the 
	first fading-null frequency as a function of fiber length, revealing HCF's $\sim$6.8$\times$ 
	reach extension relative to SMF. For SMF with $D = 17$~ps/(nm$\cdot$km), the first fading null at 40~km occurs
	at approximately 9.6~GHz, completely within the signal bandwidth required for
	100G PAM4 at 53~GBaud ($\sim$28~GHz). Even at 10~km, the null falls at only
	$\sim$19~GHz. The resulting power-fading penalty
	renders high-rate IMDD transmission infeasible at metro distances without
	external dispersion compensation.
	
	HCF fundamentally changes this constraint. With
	$D = 2$--4~ps/(nm$\cdot$km), the first fading null at 40~km shifts to
	20--28~GHz, comparable to or above the half-baud bandwidth of 100G PAM4.
	At 10~km---the critical distance for campus DCI---the null shifts from
	$\sim$19~GHz (SMF) to 40--56~GHz (HCF), comfortably accommodating 400G and
	800G transceivers. Since the first fading null frequency scales as
	$f_{\text{null}} \propto 1/\sqrt{D \cdot L}$, the dispersion-limited reach
	for a given null frequency threshold extends by the ratio
	$D_{\text{SMF}}/D_{\text{HCF}} \approx 4$--8$\times$. Links in the
	10--40~km range that require coherent transceivers or dispersion-compensating
	fiber (DCF) with SMF can use low-cost IMDD transceivers with HCF.
	
	A secondary advantage relevant to WDM systems is that HCF exhibits
	relatively flat dispersion across the C-band. In SMF, the dispersion slope
	($\sim$0.06~ps/(nm$^2\cdot$km)) causes outer WDM channels to experience
	materially different CD than the center channel, requiring per-channel
	equalization tuning. In HCF, the low absolute dispersion and near-flat
	dispersion profile across $>$100~nm of usable bandwidth allow uniform
	equalizer configurations across all WDM channels, simplifying system design
	for DWDM IMDD.

	\subsubsection*{O-Band Operation in SMF}
	It is instructive to compare C-band HCF with O-band operation in standard
	SMF, since both offer reduced dispersion relative to C-band SMF.
	Standard G.652 SMF exhibits its zero-dispersion wavelength (ZDW) near
	1310~nm, yielding $D \approx 1$--3~ps/(nm$\cdot$km) in the O-band
	(1260--1360~nm). This near-zero dispersion has motivated widespread adoption of O-band IMDD
	for short-reach standards such as IEEE~802.3bs 400G-DR4 and emerging
	800G-DR8, where links are typically under 2~km.

	However, O-band SMF carries two significant penalties that limit its
	viability at extended reach. First, fiber attenuation in the O-band is
	approximately 0.32--0.35~dB/km, compared to $\sim$0.20~dB/km at 1550~nm,
	substantially reducing the available power budget for unamplified links.
	Second, mature wideband optical amplification is absent in the O-band:
	erbium-doped fiber amplifiers (EDFAs) operate exclusively in the C- and
	L-bands, precluding amplified O-band IMDD spans. Bismuth-doped amplifiers
	covering the O-band have been demonstrated in research but are not yet
	commercially deployed. Consequently, O-band IMDD is confined to short
	unamplified links of a few kilometers at most.

	HCF at C-band uniquely combines O-band-like low dispersion
	($D \approx 2$--4~ps/(nm$\cdot$km)) with C-band attenuation ($<$0.15~dB/km
	in production fiber, approaching 0.091~dB/km~\cite{slavik_ofc2026}),
	EDFA compatibility for amplified spans, near-zero nonlinearity, and 31\%
	latency reduction. This combination extends dispersion-tolerant IMDD
	viability from the $\sim$2~km regime accessible with O-band SMF to
	metro-scale links of 10--80~km---a qualitative shift in what low-cost IMDD
	technology can address.

\subsubsection*{Comparison to G.655} Our comparisons are framed against G.652 SMF, whereas G.655 NZDSF exhibits $D \approx 3$--6~ps/(nm$\cdot$km)---numerically close to HCF's 2--4~ps/(nm$\cdot$km). On dispersion-limited IMDD reach alone, NZDSF therefore offers a comparable first-null frequency to HCF. However, G.655 was specified for long-haul DWDM with EDFA amplification and is not widely deployed in data-center, campus, or fronthaul environments, where G.652.D dominates. More importantly, HCF's other advantages over G.652---the $\sim$1{,}000$\times$ lower nonlinear coefficient, the 31\% latency reduction from near-unity group index, and the much lower thermal sensitivity of group delay---are physical consequences of air-core propagation and are preserved relative to G.655 as well.

The same comparison extends to G.657.A1 bend-insensitive fiber, increasingly
used in FTTH and urban access: it shares the chromatic dispersion of G.652.D
($\sim$17~ps/(nm$\cdot$km)), so HCF's dispersion advantage is essentially
unchanged relative to G.657.A1 plant.

Beyond chromatic dispersion, IM/DD systems face a qualitatively different
impairment from polarization-mode dispersion (PMD). While CD-induced fading
produces deterministic, frequency-periodic nulls, PMD introduces random,
time-varying spectral fading: the differential group delay (DGD) between the
two principal states of polarization fluctuates with temperature and mechanical
stress following a Maxwellian distribution, producing
unpredictable fading nulls across the signal band. Unlike coherent
receivers---which recover both field quadratures and apply MIMO-DSP to equalize
PMD---IM/DD receivers detect only total optical intensity and have no practical
mechanism to compensate for polarization-dependent fading. At 100\,GBaud, the
maximum allowable DGD is approximately 0.7--1.0\,ps for a sub-1\,dB worst-case
penalty; by contrast, standard G.652 SMF with a cabled
$\text{PMD}_\text{Q}$ of 0.20--0.26\,ps\,km$^{-1/2}$ accumulates 3--5\,ps of
DGD at metro reaches, limiting error-free IM/DD operation to roughly
20\,km. Fiber spinning during draw has
reduced the PMD coefficient of cabled DNANF to
0.046\,ps\,km$^{-1/2}$---well below the G.652 SMF
link design value of 0.1\,ps\,km$^{-1/2}$---substantially mitigating
PMD-induced fading for high-baud-rate IM/DD links.
	
	\subsection{Nonlinearity Advantage and Launch Power Budget}
	
	In SMF, the Kerr nonlinearity---arising from the intensity-dependent refractive
	index of silica---constrains practical launch power to approximately +3~dBm.
	At higher powers, self-phase modulation (SPM), cross-phase modulation (XPM),
	and four-wave mixing (FWM) introduce distortions that degrade signal quality.
	In WDM systems, XPM and FWM between co-propagating channels impose additional
	constraints on channel spacing and per-channel power, limiting aggregate
	system capacity.
	
	In HCF, the nonlinear coefficient of the air core is approximately
	1,000$\times$ lower than silica, permitting launch powers of +10 to +20~dBm
	without significant nonlinear degradation. For the unamplified IMDD links
	analyzed in this paper, the practical upper bound is set by receiver overload
	rather than fiber nonlinearity (PAM4 receiver specifications limit average
	received power to roughly +3.5~dBm per lane); amplified long-haul HCF
	experiments have demonstrated +34.5~dBm aggregate launch without nonlinear
	penalty, confirming that the fiber itself imposes no practical power ceiling.
	The stimulated Brillouin scattering
	(SBS) threshold is similarly elevated, as the acoustic phonon interaction in
	air is negligible compared to silica. For unamplified IMDD links---the
	standard configuration for DCI and fronthaul---this yields 7--17~dB of
	additional optical power budget. For WDM systems, the negligible XPM and FWM
	permit tighter channel spacing and higher per-channel launch power, increasing
	aggregate capacity without inter-channel penalty.
	
	The practical impact is quantified through a representative link budget.
	Consider a 40~km unamplified IMDD link with a conservative HCF attenuation
	of 0.17~dB/km (representative of early-generation cabled fiber; current
	production HCF achieves 0.10--0.15~dB/km):
	the total fiber loss is 6.8~dB. With a launch power of +10~dBm (conservative
	for HCF) and a PAM4 receiver sensitivity of $-$20~dBm, the available margin
	is 23.2~dB (without splice losses), comfortably accommodating splice losses (typically 4 HCF-SMF
	transitions using conservative field estimates of 0.3--0.5~dB each with optimized mode-field adapters), connector
	losses, and system margin. By comparison, the same 40~km link on SMF (at 0.2~dB/km, yielding 8.0~dB
	fiber loss) is constrained to +3~dBm launch, yielding only 15.0~dB
	margin---an 8.2~dB deficit that reflects both the higher permissible launch
	power and the lower attenuation of HCF. With higher HCF launch powers of +15~dBm, the received power reaches approximately +6.6~dBm after accounting for fiber loss (6.8~dB) and four
	HCF-SMF transitions (1.6~dB total), representing an approximately 12~dB
	improvement over an equivalent SMF link at +3~dBm launch with 8.0~dB fiber
	loss. This margin can extend reach,
	compensate for HCF-to-SMF transition losses, support higher-order modulation
	formats (PAM8), or relax receiver sensitivity requirements.
	
	\subsection{Latency Reduction}
	
	HCF exhibits a group index $n_g \approx 1.003$, compared to $n_g \approx
	1.468$ for SMF. The resulting propagation delay difference is approximately
	1.55~$\mu$s/km, yielding a 31\% latency reduction. Over a
	10~km campus link, this corresponds to approximately 15.5~$\mu$s one-way
	savings.
	
	This reduction is significant for several latency-sensitive applications. In
	5G fronthaul with URLLC requirements, the 3GPP specification mandates
	sub-millisecond end-to-end latency, of which fiber propagation delay
	constitutes a substantial fraction. For a 20~km fronthaul link, HCF saves
	approximately 31~$\mu$s one-way (62~$\mu$s round-trip), directly expanding
	the processing time available at the baseband unit. In high-frequency trading,
	HCF deployment has been justified since 2021-2022 on the basis of
	microsecond-level latency advantages, where the round-trip savings over a
	40~km link amount to 124~$\mu$s. For distributed AI training workloads
	requiring gradient synchronization across GPU clusters, reduced and
	deterministic latency improves throughput by minimizing the idle time during
	all-reduce operations across campus-scale interconnects.
	
	HCF's latency also exhibits substantially lower thermal sensitivity than SMF.
	In SMF, thermal expansion and thermo-optic effects produce latency variations of approximately
	40~ps/km/$^\circ$C. Over a 10~km link with a 20$^\circ$C diurnal temperature swing, this produces
	8~ns of latency variation---non-negligible for precision timing applications. Bare HCF reduces this
	to approximately 0.4~ns over the same link (${\sim}2$~ps/km/$^\circ$C), a $20{\times}$ improvement;
	production-coated HCF exhibits ${\sim}6$~ps/km/$^\circ$C, still representing an approximately
	$7{\times}$ improvement over SMF, providing substantially more stable and deterministic propagation
	delay. 
	The consequence extends to clock recovery. In continuous-mode IM/DD links, AR-HCF's reduced thermal delay drift proportionally slows the phase drift the clock-and-data recovery (CDR) circuit must track, permitting narrower CDR loop bandwidths and reduced jitter contribution. In optically switched and burst-mode systems, the stable propagation delay has been exploited to achieve sub-nanosecond (\(<625~\mathrm{ps}\)) clock recovery on \(60~\mathrm{ns}\) packets over hollow-core fiber without active clock-phase tracking---a function that requires continuous phase updates over SMF. 
	
	\section{Inter-Modal Interference in HCF-Based IMDD}
	
	While HCF provides substantial advantages for IMDD, inter-modal interference
	(IMI) constitutes the dominant impairment unique to HCF-based systems and must
	be carefully managed in link design. Phenomenologically, because the delayed
	higher-order modes beat against the fundamental mode at the square-law detector,
	IMI manifests mathematically and systemically as multipath interference (MPI). Therefore, from a system-level performance perspective, we will hereafter refer to this impairment as MPI.
	
	At each splice, connector, or structural perturbation, a fraction of the
	fundamental mode (LP$_{01}$) power couples into higher-order modes---primarily
	LP$_{11}$---that propagate at a different group velocity. The coupling
	coefficient per splice is typically $\kappa \approx -35$~dB for optimized
	HCF-SMF splices, though it can be significantly higher ($-$20 to $-$25~dB)
	for unoptimized connections or connectors. The MFD mismatch between HCF
	($\sim$18-25~$\mu$m) and SMF ($\sim$10~$\mu$m) is the primary driver of this
	inter-modal coupling. At subsequent coupling points, the scattered light
	re-enters the fundamental mode with a stochastic phase offset, producing
	intensity noise at the square-law detector. 
		\begin{figure*}[h]
		\centering
		\includegraphics[width=\textwidth]{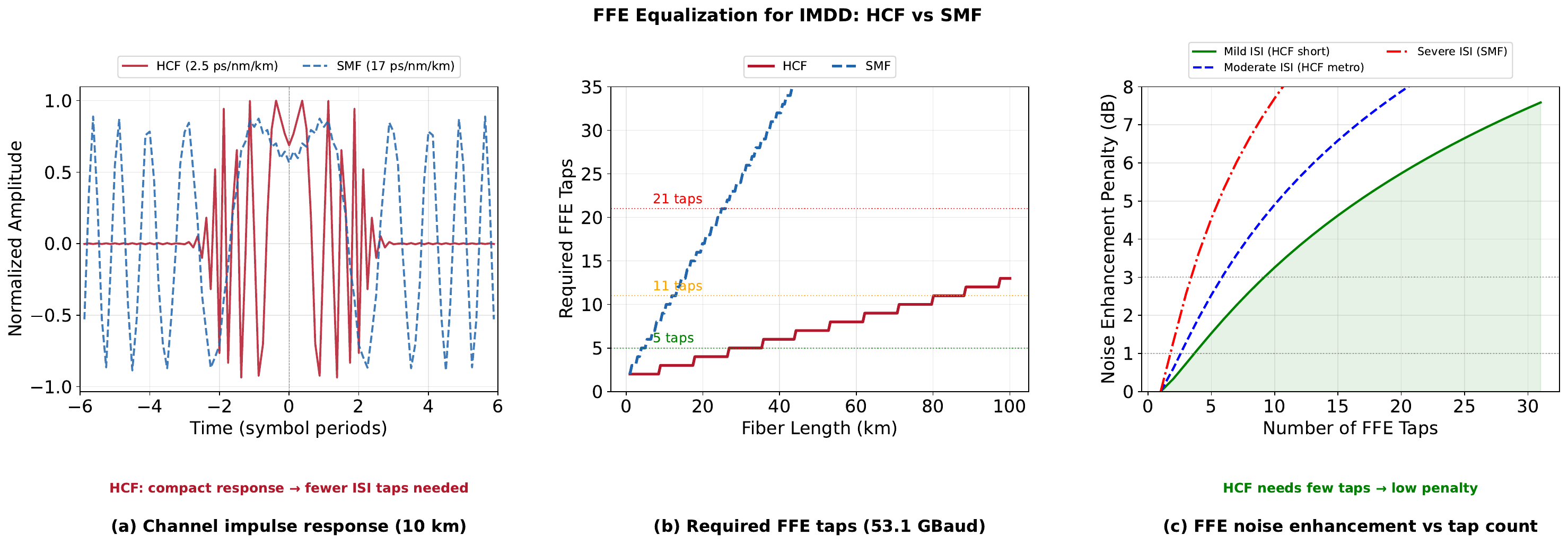}
		\caption{FFE equalization comparison for HCF versus SMF IMDD links.
			(a)~Channel impulse response at 10~km---HCF produces a compact response while
			SMF exhibits energy spreading across multiple symbol periods. (b)~Required FFE taps versus distance, showing 3--6$\times$ reduction for HCF. (c)~Noise
			enhancement penalty versus tap count. The noise enhancement curves are computed using an illustrative parametric model, and are intended to illustrate qualitative trends rather than serve as a predictive model.}
		\label{fig:ffe}
	\end{figure*}
	
	IMDD systems are fundamentally more sensitive to MPI than coherent
	systems. Coherent receivers can employ MIMO processing
	to separate and cancel interfering modes; IMDD receivers detect only the total
	intensity and cannot distinguish modal contributions. For PAM4, the MPI power
	ratio must remain below approximately $-$30~dB (relative to the signal) to
	maintain penalty below 1~dB while PAM8 requires MPI below $-$35~dB.
	
	The principal defense mechanism in HCF is differential modal attenuation (DMA).
	Higher-order modes in AR-HCF experience substantially higher propagation loss
	than the fundamental mode. In NANF designs optimized for low loss, DMA values
	on the order of 2--12~dB/km have been reported, while
	high-mode-purity designs such as the 4T-DNANF achieve DMA exceeding
	6,500~dB/km~\cite{gao2025}. With a representative DMA of 12~dB/km, the
	LP$_{11}$ mode power is attenuated by 60~dB over 5~km, reducing its
	contribution to MPI to negligible levels. For IMDD links exceeding
	approximately 5~km, the fiber's intrinsic DMA provides sufficient mode
	suppression to maintain MPI well below the $-$30~dB threshold.
		
	This leads to a distance-dependent fiber design trade-off that can be
	illustrated with two representative fiber designs. A low-loss fiber (type~A)
	with 0.1~dB/km fundamental-mode attenuation and DMA of 8~dB/km provides only
	16~dB of higher-order mode suppression at 2~km---insufficient for PAM4, which
	can incur $>$2~dB OSNR penalty at this MPI level. A high-DMA fiber (type~B)
	with 0.5~dB/km fundamental-mode loss but DMA of 25~dB/km provides 50~dB
	suppression at the same 2~km distance, rendering MPI negligible. Short links
	($<$5~km, typical for intra-DC) therefore benefit from fiber designs with high
	DMA ($>$15~dB/km), even at the expense of higher fundamental-mode loss. Longer
	links (10--80~km) can use fibers optimized for minimum fundamental-mode loss,
	as the propagation distance provides adequate DMA-based suppression. 
	
	Recent fiber designs are relaxing this trade-off. The 4T-DNANF achieves
	0.1~dB/km fundamental-mode loss with DMA of 430~dB/km, and in a variant
	optimized for modal purity, 0.13~dB/km with DMA exceeding
	6,500~dB/km~\cite{gao2025}---effectively eliminating MPI at any practical link
	length. Splice engineering further mitigates MPI at the source, achieving inter-modal coupling below $-$35~dB per
	splice~\cite{slavik_ofc2026}, minimizing the initial excitation of higher-order
	modes. With current fiber and splicing technology, MPI is a managed impairment
	for links exceeding $\sim$5~km, and ongoing advances are extending this to
	shorter links.

	\section{Equalization Benefits}
	
	Beyond MPI management, the reduced dispersion of HCF directly benefits the equalization requirements at the receiver, as analyzed next. Feed-forward equalization (FFE) is the standard technique for mitigating
	intersymbol interference (ISI) in IMDD receivers. Figure~\ref{fig:ffe} provides
	a comprehensive comparison of HCF and SMF equalization requirements: Fig.~\ref{fig:ffe}(a) 
	shows the channel impulse response at 10~km, Fig.~\ref{fig:ffe}(b) quantifies the required 
	FFE tap count as a function of distance, and Fig.~\ref{fig:ffe}(c) shows how noise enhancement 
	penalty grows with tap count under mild, moderate, and severe ISI conditions. The FFE operates as a
	finite impulse response (FIR) filter with $N$ taps, each spaced at the symbol
	period. While additional taps improve ISI cancellation, they simultaneously
	amplify receiver noise---the noise enhancement penalty (NEP)---which represents the
	fundamental performance limit of linear equalization. The noise enhancement curves in Fig.~\ref{fig:ffe}(c) are computed using an illustrative parametric model of the form $\text{NEP} = 10\log_{10}[1 + \alpha(N-1)^{1.2}]$, with coefficients $\alpha = 0.08$, $0.15$, and $0.35$ representing mild, moderate, and severe ISI conditions respectively. These coefficients define a phenomenological model intended to illustrate the qualitative trends of noise enhancement rather than serve as a predictive model, consistent with the severe equalization limitations of standard SMF reported in recent IMDD literature~\cite{yang2025}.
	
	
	The ISI memory length, and hence the required number of FFE taps, scales
	linearly with the accumulated dispersion $D \cdot L$. Specifically, the
	minimum tap count can be estimated as
	$N_\text{taps} \approx \lceil 2 D L \Delta\lambda / T_\text{sym} \rceil + 1$,
	where $\Delta\lambda = \lambda^2 B / c$ is the signal spectral width, $B$ is
	the baud rate, and $T_\text{sym}$ is the symbol period. Since HCF dispersion
	is approximately 7$\times$ lower than SMF, the ISI memory is correspondingly
	reduced. At 10~km and 53~GBaud, SMF requires approximately 9 T-spaced FFE taps while HCF requires only 3---a 3$\times$ reduction at 10~km, increasing to 5--6$\times$ at metro distances. Taps may be
	symbol-spaced ($T$-spaced) or fractionally spaced ($T/2$-spaced); the latter
	provides improved performance at the cost of doubled computational load, but
	HCF's reduced tap count makes fractionally-spaced implementations practical
	even at 106~GBaud for 800G systems.

	This reduction has three direct consequences. First, the noise enhancement
	penalty decreases substantially: a 5-tap FFE under mild ISI conditions (typical for HCF) incurs $<$2~dB of
	noise enhancement, while a 15-tap FFE can incur 3--10~dB depending on ISI severity (3--5~dB for HCF metro conditions, up to~$\sim$10~dB for severe SMF-like ISI), eroding much of the
	equalization gain. Second, the computational load scales linearly with tap
	count---a 5-tap FFE requires 4$\times$ fewer multiply-accumulate operations
	per second than a 21-tap FFE, directly reducing DSP power consumption within
	the thermal constraints of QSFP-DD and OSFP form factors. Third, the reduced
	equalizer complexity enables smaller transceiver ASICs with lower die cost,
	supporting the ``light-DSP'' IMDD transceiver paradigm that targets coherent-like
	reach with direct-detect cost.

	Beyond linear FFE, decision-feedback equalizers (DFE) and nonlinear
	equalizers such as Volterra series filters are commonly employed in
	high-performance IMDD receivers. However, the reduced ISI in HCF channels
	diminishes the marginal benefit of these higher-complexity equalizers. DFE,
	which uses past decisions to cancel post-cursor ISI, provides minimal
	additional gain when the channel impulse response is compact. Volterra
	equalizers, which compensate for signal-signal beating products from the
	square-law detector, show reduced improvement when the frequency-dependent
	fading is minimal. Consequently, HCF links can achieve near-optimal
	performance with FFE alone, whereas SMF links at equivalent distances
	typically require FFE+DFE or Tomlinson--Harashima precoding (THP) at the
	transmitter.
	
	Experimental results confirm these benefits. Transmission beyond 200~Gb/s PAM-4 over 20~km of NANF has been demonstrated using THP combined with a modified multi-output FFE, reducing computational processing time by about 64.6\% compared to a conventional FFE. Real-time 800G transmission over 5~km of HCF has also been achieved with FPGA-based FFE/DFE at 106~GBaud, showing that reduced equalizer complexity can be implemented in practical systems. In addition, 100~Gb/s PAM-4 PON transmission over 20~km ofAR- HCF has been demonstrated using low-complexity DSP, achieving a 42.5~dB power budget with simplified equalization.

	\section{Application Scenarios and Deployment Economics}
	
	\subsection{Application Scenario Analysis}
	
	Five application scenarios span the IMDD deployment landscape, each with
	distinct distance requirements, cost sensitivity, and HCF value drivers
	(Table~\ref{tab:scenarios}).
	
	\textbf{Intra-data center ($<$2~km):} Fiber attenuation is negligible at
	these distances, and the primary HCF value is 31\% latency reduction for
	distributed computing synchronization, disaggregated storage, and GPU-to-GPU
	communication in AI training clusters. With link lengths typically under 500~m,
	the absolute HCF fiber cost per link is below
	\$5,000~\cite{microsoft2024}, which is comparable to the cost of a single
	high-end optical transceiver. Deployment is always
	greenfield (new rack/pod construction), minimizing the incremental cost of
	HCF. High-DMA fiber ($>$15~dB/km) is preferred to suppress MPI at these
	short distances. Yang et al.\ demonstrated 800G over 5~km HCF for this
	application~\cite{yang2025}. HCF deployment is justified at current pricing
	for latency-sensitive workloads, particularly in AI training clusters where
	gradient synchronization latency directly impacts training throughput.
	
	\textbf{Campus DCI (2--10~km):} Both latency reduction and low CD provide
	value. HCF enables 400G--800G per wavelength using IMDD at distances where
	SMF would require coherent transceivers or would incur significant dispersion
	penalty for 400G rates. Deployment is typically greenfield, and the moderate
	distances limit absolute HCF fiber cost to approximately
	\$25,000--50,000 for a 5~km link~\cite{microsoft2024}. Cost-competitiveness for hyperscalers is projected during
	2025--2027.
	
	\textbf{Metro DCI (10--80~km):} Low dispersion is the dominant advantage,
	as SMF power fading becomes the binding constraint beyond $\sim$10~km. SMF
	at 40~km requires either coherent transceivers or dispersion-compensating
	modules for PAM4 IMDD. HCF eliminates both requirements, enabling the same
	low-cost IMDD transceivers used for intra-DC to operate at metro distances. Low-loss
	fiber designs ($<$0.15~dB/km) are preferred at these distances to minimize
	accumulated attenuation, while the longer propagation length ensures adequate
	DMA-based MPI suppression. For a 40-channel DWDM system, replacing coherent
	transceivers with IMDD modules saves approximately
	\$1,000--2,000 per transceiver, yielding aggregate
	savings exceeding \$100,000 per link that can offset the HCF fiber premium at
	current pricing for links up to approximately 20~km. 
	
	Early hyperscaler adoption confirms the commercial trajectory. AWS has
	confirmed production deployment of HCF at approximately five to ten
	availability-zone interconnect sites as of early 2026, motivated primarily
	by latency reduction and reach extension rather than transceiver
	cost. Microsoft Azure has deployed
	$>$1{,}280~km of live HCF infrastructure, reporting 0.091~dB/km operational
	loss and zero field failures. These deployments are
	latency-driven---exploiting the $\sim$31\% reduction in propagation delay
	relative to SMF---and do not yet confirm large-scale IMDD adoption at metro
	distances. Supply constraints currently limit deployment volume, with
	manufacturers targeting expansion from tens of thousands to hundreds of
	thousands of kilometers of annual production capacity to meet hyperscaler
	demand. Broad HCF deployment for cost-motivated
	metro DCI---where the IMDD transceiver savings described above become the
	primary economic driver---is projected during 2027--2030, contingent on
	fiber cost reduction and manufacturing scale-up.
	
	\textbf{5G fronthaul ($<$20~km):} Fronthaul links connecting remote radio units (RRUs) to baseband units (BBUs) require 25--100~Gb/s capacity, low latency for URLLC compliance, and low cost at scale. HCF addresses all three requirements, and full-duplex DWDM IMDD over AR-HCF has been demonstrated. However, fronthaul is mainly brownfield: where deployed as a transport segment in centralized-RAN architectures, 5G fronthaul runs largely over installed G.652 SMF---operator surveys attribute roughly 59\% of the fronthaul connectivity mix to WDM or dark fiber---carrying CPRI/eCPRI over SFP28/SFP56 grey or CWDM/DWDM modules~\mbox{\cite{heavy_reading2020}}. Full fiber replacement for HCF upgrades would therefore incur the civil-works burden.
	
	\textbf{PON ($<$40~km):} 100~Gb/s PAM-4 PON over 20~km of AR-HCF
	with a 42.5~dB link budget has been demonstrated in the literature (using an amplified ONU
	receiver), confirming technical feasibility for
	next-generation access. An additional advantage of HCF for PON is its very low backscattering, measured more than 40\,dB below that of silica-core fiber. Conventional PON systems separate upstream and downstream wavelengths to suppress Rayleigh-backscatter and reflection crosstalk in single-fiber bidirectional operation. Recent AR-HCF experiments have demonstrated co-frequency full-duplex IM/DD with only about 0.5\,dB backward-crosstalk penalty, suggesting that HCF could enable tighter wavelength plans and simpler bidirectional designs, although full PON-level validation remains to be shown.

	Unlike point-to-point links, PON is inherently point-to-multipoint:
		a single OLT serves 32--64~ONUs (up to 128 in XGS-PON) via passive splitters,
		so feeder-fiber cost is amortized across many subscribers.
		While this reduces the absolute impact of HCF cost, the per-subscriber
		cost remains sensitive to the feeder-fiber price, since any premium
		applies directly to the shared feeder segment, whereas distribution
		and drop fibers remain SMF.
		Accordingly, the PON results in Fig.~\ref{fig:economics} model a 20~km
		feeder segment---the baseline maximum fiber distance standardized across
		ITU-T PON generations---and should be read as an upper bound on HCF's
		per-subscriber cost impact, which is lower for the shorter feeders common
		in practice.
	Given the large installed base of SMF and strong cost sensitivity
	in access networks, HCF adoption in PON is unlikely without
	near-parity in fiber cost.
	
	\mbox{\textbf{Fixed-mobile access convergence.}} Where a single fiber plant serves both residential PON and mobile fronthaul/backhaul, the incremental civil-works cost is shared across services and the absolute HCF--SMF fiber differential is spread over a larger revenue base, marginally improving HCF's case. Two factors still constrain near-term HCF penetration in converged access, however: (i) the passive splitter architecture of PON multiplies every HCF fiber meter by the branch count, and (ii) the dominant cost saving in converged builds comes from sharing ducts with SMF rather than replacing them with HCF. Convergence is therefore likely to accelerate HCF in \mbox{\emph{converged metro backhaul}} (20--80~km aggregation segments) well before it changes access/PON economics.

\subsection{Power-over-Fiber and IMDD Co-Propagation}
AR-HCF enables converged links in which a high-power optical beam for power-over-fiber (PoF) co-propagates with data in the same fiber—difficult to realize at useful power levels in silica SMF due to stimulated Raman scattering (SRS) and Kerr-induced nonlinear coupling. In AR-HCF, the guided mode overlaps the glass at the tens-of-parts-per-million level, reducing effective nonlinearity by $\sim$10$^{3}$ relative to SMF and strongly suppressing SRS-induced noise transfer. Altuna et~al.\ confirmed this by comparing 5G~NR ARoF transmission with co-propagating PoF over 11.1~km AR-HCF, 10~km multicore fiber (MCF), and 9.8~km SMF, observing negligible degradation in HCF and MCF but severe penalties in SMF~\mbox{\cite{altuna_lpor_2024}}. Recent demonstrations include 10~W CW 1064~nm co-propagated with a 4.64~Tbit/s bidirectional C-band IMDD link over 3~km AR-HCF at 24.7\% optical power transmission efficiency (OPTE), system energy efficiencies of 9.9\% over 3.1~km and 0.9\% over 11.1~km AR-HCF while carrying 5G~NR ARoF~\mbox{\cite{altuna_lpor_2024}}, and simultaneous C/L-band power and data delivery over 3.1~km NANF. A recent review places these results in the broader B5G radio-power-over-fiber fronthaul context.

HCF core diameter presents a fundamental trade-off: a larger air core reduces confinement loss ($\propto(\lambda/R)^{4}$) and optical intensity at high launch power, yet increases inter-modal interference
and incurs mode-field adapter losses at HCF$\leftrightarrow$SMF transitions. This favors a moderately larger core (${\sim}28$--$30\,\mu$m) HCF with high differential modal attenuation for PoF application.

\mbox{\textbf{Multiplexing strategy for high-power PoF.}} Three approaches separate pump and data. First, AR-HCF's multi-window transmission enables spectral isolation between $\sim$1~$\mu$m PoF (e.g., 1064~nm) and C-band data via the intervening anti-resonance bands, which act as high-loss barriers. Second, both channels may share a common low-loss window and be separated by coarse WDM filtering, at the cost of reduced intrinsic isolation. Third, in the tens-of-watts regime and beyond, hermetic end-sealing and high-power-rated connectors are required to mitigate thermal and interface-damage effects at splice and termination points. In practice, the co-propagating power limit is set by connectors, splices, and photoconversion efficiency at the endpoints rather than by the fiber itself: kilowatt-class single-mode delivery over kilometer-scale AR-HCF has been demonstrated in dedicated power-delivery configurations.

\mbox{\textbf{Comparison with multicore fiber.}} MCF provides an alternative using spatial rather than spectral separation, with pump and data carried in different cores and inter-core crosstalk typically below $-30$~dB over practical link lengths. As shown in~\mbox{\cite{altuna_lpor_2024}}, MCF similarly suppresses SRS-induced noise transfer in the ARoF band, and at comparable lengths the PoF system energy efficiencies of MCF and AR-HCF are broadly similar (1.5\% at 10~km MCF versus 0.9\% at 11.1~km AR-HCF). The practical differentiator is therefore form factor and termination: AR-HCF retains a standard fiber outer diameter (125--250~$\mu$m) compatible with conventional cabling and can be fusion-spliced or interconnected to SMF with modest loss, whereas MCF requires fan-in/fan-out devices at every termination. In fairness, HCF PoF likewise requires WDM mux/demux couplers ($\sim$0.3--0.5~dB each) to combine and separate the pump and data, comparable in magnitude to MCF fan-in/fan-out losses ($\sim$0.5--1~dB each); the decisive advantage of both over SMF is SRS suppression rather than termination loss. HCF nonetheless favors deployment simplicity through its standard outer diameter and direct SMF splicing.
Overall, AR-HCF uniquely enables single-core spectral co-propagation of high-power PoF with C/L-band IMDD data, supporting access and fronthaul links where a single fiber can deliver both power and data to remote units.

	\begin{table}[!t]
		\centering
		\caption{HCF-IMDD Application Scenario Summary. TBD: to be determined}
		\label{tab:scenarios}
		\renewcommand{\arraystretch}{1.15}
		\begin{tabular}{@{}p{1.3cm}p{0.7cm}p{1.6cm}p{1.3cm}p{1.1cm}@{}}
			\toprule
			\textbf{Scenario} & \textbf{Distance} & \textbf{Key HCF Benefit} &
			\textbf{Fiber Priority} & \textbf{When} \\
			\midrule
			Intra-DC  & $<$2\,km  & Latency ($-$31\%) & High DMA & Now \\
			Campus DCI & 2--10\,km  & Latency + low CD & High DMA & 2025--27 \\
			Metro DCI & 10--80\,km & Low CD (7$\times$) & Low loss & 2027--30 \\
			5G FH     & $<$20\,km & Latency + capacity & Balanced & 2030+ \\
			PON       & $<$40\,km & Low CD, reach & Low loss & TBD \\
			\bottomrule
		\end{tabular}
	\end{table}

	\subsection{Deployment Cost Analysis}
	\begin{figure*}[!t]
		\centering
		\includegraphics[width=\textwidth]{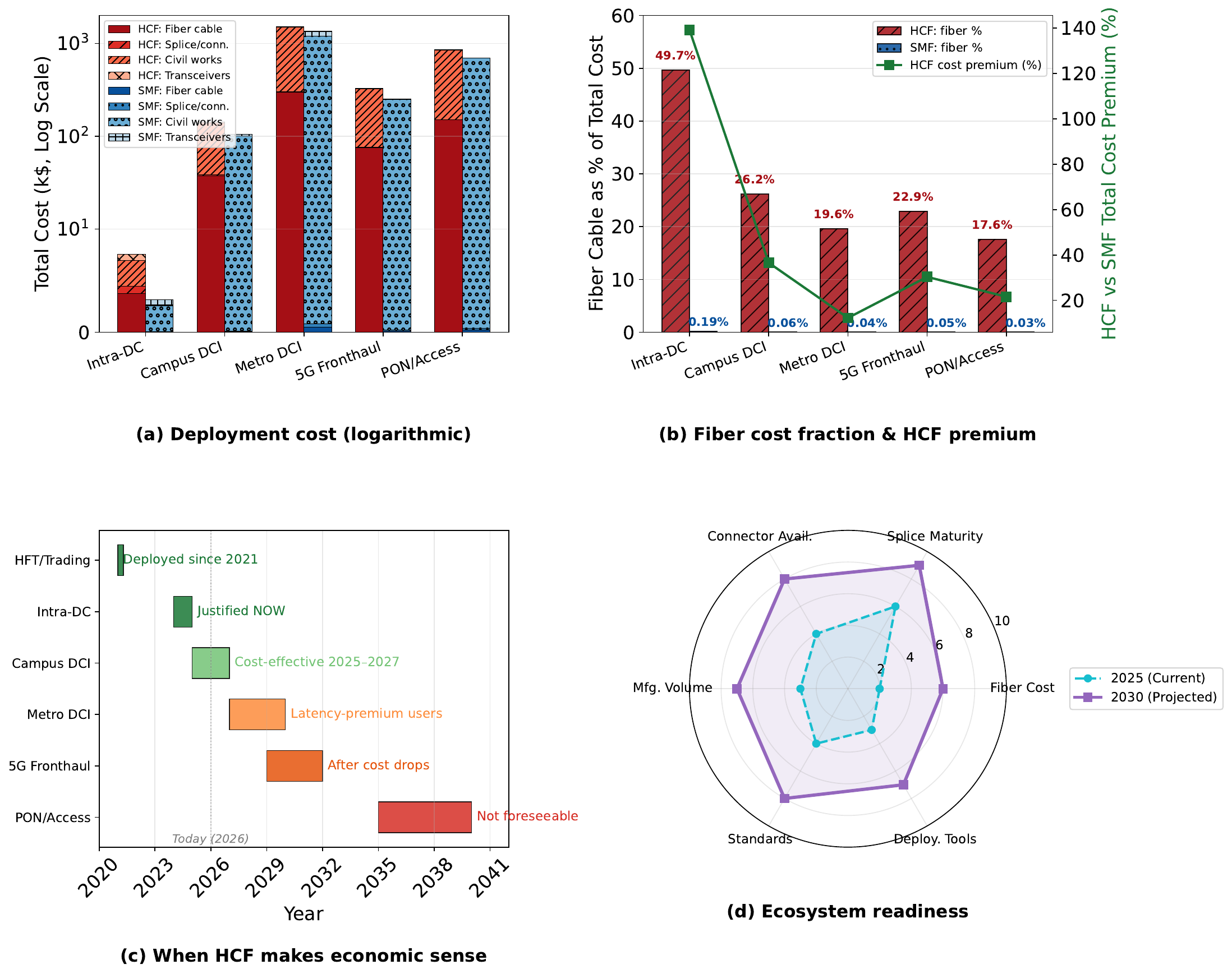}
		\caption{Deployment economics and adoption outlook. (a)~Total cost breakdown
			for HCF (warm colors, left bars) versus SMF (cool colors, right bars) across
			five scenarios on a logarithmic scale, showing that civil works dominate
			outside-plant costs. (b)~Fiber-cable cost as a percentage of total deployment
			(bars) and the HCF total-cost premium over SMF (line), showing the fiber
			premium has diminishing impact as civil works grow. (c)~Adoption timeline
			indicating when each scenario becomes economically justified; the
			vertical-axis label ``HFT/Trading'' denotes high-frequency trading.
			(d)~Ecosystem-readiness radar comparing 2025 with projected 2030 maturity
			across fiber cost, splice maturity, connector availability, manufacturing
			volume, standards, and deployment tools. Civil-works costs are representative
			estimates that vary with geography, urban density, and permitting; metro DCI
			is modeled as a 40-channel DWDM system (80~transceivers) and campus DCI as
			8-channel (16~transceivers), with the remaining scenarios as single
			point-to-point channels. Ecosystem scores (0--10) are qualitative author
			assessments based on published literature and industry announcements as of
			early 2026.}
		\label{fig:economics}
	\end{figure*}
		
		The most frequently cited barrier to HCF adoption is the fiber cost premium. At approximately \$5,000--10,000/km for low-count production AR-HCF cables (2024--2025)~\mbox{\cite{mapyourtech2026}} versus roughly \$10--15/km for standard G.652.D SMF in bulk procurement~\mbox{\cite{mapyourtech2026}}, the raw fiber-price ratio appears prohibitive. However, this metric is misleading when considered in the context of total deployment cost (Fig.~\ref{fig:economics}).
	
\mbox{\textbf{Cost model methodology and data sources.}} The cost figures used in this section are drawn from published infrastructure-economics literature, and vendor and standards data. SMF cable pricing (\$10--15/km for loose-tube G.652.D, 24-fiber count) is consistent with reported bulk-procurement ranges in the techno-economic literature~\mbox{\cite{yingfeng_2026}}. HCF pricing (\$5,000--10,000/km for production AR-HCF cables, 2024--2025) is based on reported commercial estimates~\mbox{\cite{mapyourtech2026}}; the authors note that HCF list prices are not publicly disclosed by most vendors and that this range represents analyst estimates rather than audited procurement data. Coherent transceiver pricing (\$1,500--2,500 for 400ZR/400ZR+ QSFP-DD) and IMDD transceiver pricing (\$200--400 for 400G DR4/FR4) are representative market values for 2024--2025~\mbox{\cite{lightcounting2024}}. Typical 400ZR module power (15--18~W) and IMDD DR4 power (7--12~W) are drawn from the OIF 400ZR Implementation Agreement~\mbox{\cite{oif2020}}. Outside-plant civil-works costs (trenching, ducting, permitting, restoration, and labor) are informed by the survey in~\mbox{\cite{nyarkoboateng2020}} and the European Commission's broadband deployment cost review~\mbox{\cite{ec_gigabit_2023}}, which together support a representative range of \$5,000--35,000/km depending on urban density and substrate (rural soft-dig through urban hard-dig). The \$5,000/km estimate used for intra-DC/campus and \$15,000--35,000/km for outside-plant scenarios are author-derived midpoints from this range and will vary with geography and regulatory environment.
	
	\mbox{\textbf{CAPEX and OPEX decomposition.}} CAPEX comprises four components: (1)~\mbox{\emph{fiber cable}}---cable cost per route-kilometer; (2)~\mbox{\emph{splices and connectors}}---at approximately \$150 per HCF splice and \$200 per HCF connector (our estimate) versus \$15 and \$10 for SMF ~\mbox{\cite{splice_2026}}, with one splice per 2~km and an HCF-SMF transition pair at each link end; (3)~\mbox{\emph{civil works}}---the dominant outside-plant line item; and (4)~\mbox{\emph{transceivers}}---two transceivers per link end at the pricing ranges above. OPEX comprises: (1)~\mbox{\emph{transceiver power}}---the differential between coherent 400ZR ($\sim$16.5~W) and IMDD DR4 ($\sim$9.5~W) is $\sim$7~W per module; at \$0.10/kWh over a 10-year lifecycle this represents approximately \$61 per module, or roughly \$4,900 for a 40-channel DWDM link, a modest addition to the transceiver-price differential; (2)~\mbox{\emph{maintenance labor}}---periodic monitoring sweeps, repair dispatches, and splice re-certifications; (3)~\mbox{\emph{personnel training}}---one-time training of splicing technicians on HCF-specific fusion procedures; and (4)~\mbox{\emph{repair parts and consumables}}---HCF-specific splice sleeves, mode-field adapters, and connectors. On a 10-year discounted basis, OPEX typically represents 15--25\% of total cost of ownership for outside-plant deployments~\mbox{\cite{nyarkoboateng2020}}, and the CAPEX-driven conclusions in Fig.~\ref{fig:economics} remain qualitatively unchanged once OPEX is included.

	The assumed $\sim$10$\times$ HCF-to-SMF splice-cost ratio is an authors'
	estimate rather than a measured figure: it reflects current specialty-splicer
	cost, longer process time ($\sim$100~s per splice versus $\sim$35~s for SMF),
	operator training, and HCF-specific consumables such as mode-field adapters
	and hermetic protection. As dedicated HCF splicers and operator training
	mature, this ratio should narrow.
	
	In outside-plant deployments (metro, fronthaul, PON), civil works---trenching,
	ducting, permitting, and surface restoration and labor---account for 60--80\% of total
	cost~\cite{ec_gigabit_2023}. In high-cost urban deployments (where civil works can reach \$100k--500k/km as reported in~\cite{ec_gigabit_2023}), fiber cable represents only 5--10\% of total cost. In the moderate-cost scenarios modeled here (\$25k--35k/km), fiber represents approximately 18--26\%. For a 40~km metro DCI deployment, the HCF total cost premium over SMF is approximately 12\%, as civil works dominate both budgets. In intra-DC deployments where civil works are
	minimal, fiber represents a larger share (40--60\%), but absolute distances
	are short, limiting total HCF cost per link to
	\$5,000--20,000~\cite{microsoft2024}.
	
	A critical economic factor at metro distances is coherent transceiver
	avoidance. When SMF-based IMDD links cannot operate beyond $\sim$10~km due to
	dispersion, the alternative is coherent transceivers at $\sim$\$1,500--2,500
	each versus \$200--400 for IMDD---a differential of
	approximately \$1,000--2,000 per transceiver. For a 40-channel DWDM link
	requiring transceivers at both ends, the aggregate transceiver cost
	differential can exceed \$100,000 per link. Additionally, coherent modules
	targeting the OIF 400ZR specification consume 15--18~W versus
	$\sim$7--12~W for IMDD DR4; over 40~channels and a 10-year
	lifecycle, this power differential represents substantial operational
	expenditure in electricity and cooling. These savings can offset the HCF fiber
	premium for many greenfield metro routes.

	The greenfield versus brownfield distinction is critical. In greenfield
	projects, the incremental cost of choosing HCF over SMF is limited to the
	fiber cost differential, as civil works are incurred regardless of fiber type.
	In brownfield upgrades, the full deployment cost including cable replacement
	must be justified by HCF benefits alone. Intra-DC and campus DCI are
	predominantly greenfield; 5G fronthaul and PON are predominantly brownfield.
	
	HCF manufacturing costs are declining rapidly as production volumes
	increase~\cite{microsoft2024}. With manufacturers scaling
	production, HCF is expected to follow manufacturing experience curves analogous to those observed in other photonic component industries. As production volumes increase, we estimate that unit costs could decline to approximately \$1,000--2,000/km by the end of the decade, eventually approaching parity with premium specialty solid-core fibers. The total value proposition for HCF---encompassing latency
	value, coherent avoidance savings, extended IMDD reach, and avoidance of
	future re-trenching costs---frequently justifies the current premium in
	greenfield scenarios.
	
	\mbox{\textbf{Derivation of the ecosystem-readiness scores}} The six dimensions in the radar chart are qualitative assessments scored on a 0--10 scale, with the following evidence basis for each 2025--2026 score. \mbox{\emph{Fiber cost}} (score~3): anchored in the \$5,000--10,000/km 2024--2025 price range~\mbox{\cite{mapyourtech2026}} versus a projected \$1,000--2,000/km for 2030, which is itself uncertain. \mbox{\emph{Splice maturity}} (score~6): Feng~et~al.\ report 0.043~dB mean HCF--HCF splice loss with 100\% yield across 30~trials using the Fujikura FSM-100P~\mbox{\cite{feng_ofc2026}}; Suslov et~al.\ report $<$0.2~dB HCF--SMF interconnect loss~\mbox{\cite{slavik_ofc2026}}. \mbox{\emph{Connector availability}} (score~3): commodity MPO/LC connectors for HCF are not yet available; laboratory demonstrations show 0.1--0.3~dB insertion loss but mass-manufacturable designs remain in development. \mbox{\emph{Manufacturing volume}} (score~3): based on Microsoft's disclosed deployments~\mbox{\cite{microsoft2024}} and scaling announcements from YOFC, Corning, Heraeus, and OFS; estimated global annual production is in the low tens of thousands of kilometers. \mbox{\emph{Standards}} (score~4): ITU-T SG15 is developing a technical report on HCF; CCSA has approved an HCF test-method project; IEC SC86A work is ongoing. \mbox{\emph{Deployment tools}} (score~5): Fujikura FSM-100P AR-HCF splicers are commercially available; HCF-optimized bidirectional OTDR kits from EXFO (NS-348x, $\sim$150~km range) and VIAVI launched in 2025--2026; operator training and standardized workflows are still maturing. The 2030 projections (scores 7--9) assume continued scale-up, published ITU-T standards, commodity connector availability, and maturing distributed monitoring---each of which is in early development today. These scores are intended as a communication device and should be treated as illustrative rather than predictive.
	
	\subsection{Challenges, Limitations, and Open Questions}
	\label{sec:challenges}
	
	The following open technical and economic challenges could slow or reshape the adoption trajectory described above.
	
	\mbox{\emph{Connector maturity.}} Field-deployable commodity connectors with sub-0.3~dB insertion loss are not yet available in production quantities. Laboratory demonstrations show 0.1--0.3~dB insertion loss, indicating the underlying physics are tractable; the remaining work is in mass-manufacturable ferrule, alignment-sleeve, and end-cap designs compatible with existing MPO and LC form factors. Until commodity HCF connectors are available, deployments must rely on HCF-to-SMF fusion splicing at patch-panel transitions or vendor-specific assemblies.
	
	\mbox{\emph{Modal purity under field perturbations.}} The DMA-based MPI suppression analyzed in Section~IV assumes that fiber geometry is preserved end-to-end. Mechanical stress, tight bends, or temperature cycling of the cable can deform the capillary structure and locally alter the mode-coupling coefficient, potentially raising MPI. Reliability data under the temperature-humidity and mechanical-load protocols of Telcordia GR-20 and IEC~60794 are still being accumulated for cabled production HCF.
	
	\mbox{\emph{CO\textsubscript{2} absorption and gas contamination.}} As discussed in Section~II, ambient CO\textsubscript{2} diffusing into an unsealed air core produces narrow absorption lines concentrated in the L-band, with only weaker lines in the C-band. For the C-band IMDD scenarios analyzed here they are not a first-order impairment; for L-band or extended-reach links in unsealed cables, robust end-capping or spectral mitigation is required. Current production HCF uses hermetically sealed end caps to limit ingress, though long-term diffusion behavior over 10--20~year cable lifetimes is not yet fully characterized.
	
	\mbox{\emph{Water and humidity ingress.}} A breached hollow core draws in water by capillary action, and a flooded section becomes effectively opaque at telecom wavelengths. Because trapped water cannot be expelled by splicing, a flooded segment generally needs physical replacement and hermetic re-termination rather than the simple spliced repair that would suffice on SMF---a consideration for outside-plant routes prone to flooding. These constraints favour keeping HCF hermetically sealed and controlling humidity during splicing.

	\mbox{\emph{Polarization under bending and twisting.}} Bending and twisting of AR-HCF perturb the capillary geometry and induce polarization-dependent loss and polarization mode dispersion (PMD) with a stronger dependence on curvature than in SMF. For IMDD this has no first-order effect on BER. However, if the same fiber infrastructure is later repurposed for coherent transmission or polarization-division multiplexed applications, polarization behavior will become a material constraint and should be considered at the installation planning stage.
	
	\mbox{\emph{Monitoring and fault location.}} Rayleigh backscatter in HCF is typically 15--30~dB weaker than in SMF, limiting the effectiveness of conventional single-ended OTDR measurements. Bidirectional OTDR with high dynamic range and HCF-aware post-processing, as offered by EXFO and VIAVI since 2025--2026, addresses fault location and splice characterization for link certification. Key gaps remain in distributed acoustic and temperature sensing over HCF, multi-vendor tool interoperability, and standardized operator workflows for long-term asset management.
	
	\mbox{\emph{Volume manufacturing and yield.}} 
		AR-HCF fabrication uses a stack-and-draw process in which the wall thicknesses of nested capillaries must be controlled to sub-micron (often $\sim$100~nm-scale) tolerances over kilometer-scale draws to preserve the antiresonant transmission condition. This differs fundamentally from the MCVD/OVD processes used for commodity G.652 SMF. Current production volume is on the order of thousands of kilometers per year from a small number of suppliers (e.g., Microsoft/Lumenisity, YOFC, Linfiber). Meeting hyperscaler-scale deployments on the order of 10{,}000--15{,}000~km will require substantial capacity expansion and/or new entrants, and the cost trajectory assumed in Section~VI-C depends on such scaling.
	
	\mbox{\emph{Standardization and cross-vendor compatibility.}} 
		No ITU-T Recommendation currently exists for HCF; standardization efforts are ongoing within ITU-T Study Group~15, alongside activities in CCSA and IEC SC86A. In the absence of formal standards, key parameters---including core diameter, outer cladding diameter (typically 125--250~$\mu$m depending on design), mode-field properties, hermetic sealing, and acceptance metrics for DMA, IMI, and PMD---vary between suppliers. As a result, multi-vendor deployment is not yet a drop-in exercise, and interoperability generally requires vendor-coordinated qualification. Operators planning multi-vendor sourcing should therefore anticipate dedicated testing until standards mature.

\mbox{\emph{Economic sensitivity.}} The roadmap projections reported here are sensitive to the HCF fiber-price trajectory, which depends on manufacturing volume, demand and supplier competition. If HCF prices remain above \$4,000/km through 2030, cost-driven metro DCI adoption would likely be delayed, while latency-driven intra-DC adoption would be largely unaffected. A faster-than-expected cost decline would pull forward fronthaul adoption. The \$1,000--2,000/km by 2030 projection used here is an illustrative scenario based on current manufacturing trends rather than a firm forecast.
	
	\section{Conclusions and Outlook}
	\label{conclusion}
	This paper has presented a comprehensive analysis of HCF for IMDD optical
	networks, encompassing propagation advantages, the MPI challenge, equalization
	benefits, and deployment economics.
	
	HCF addresses the three principal limitations of IMDD: chromatic dispersion of
	2--4~ps/(nm$\cdot$km) extends dispersion-limited reach by 4--8$\times$,
	pushing IMDD viability from $\sim$10~km on SMF to 40--80~km on HCF; negligible nonlinearity permits launch powers of +10 to +20~dBm, improving unamplified link
	budgets; and 31\% latency reduction with $\sim$20$\times$ lower thermal
	sensitivity provides deterministic, low-delay propagation.
	
	IMI---the dominant HCF-specific impairment for IMDD---is effectively managed
	by fiber designs with DMA exceeding 12~dB/km for links longer than $\sim$5~km.
	The 4T-DNANF design, with 0.1~dB/km loss and 430~dB/km DMA (or 0.13~dB/km
	with $>$6,500~dB/km DMA in a modal-purity-optimized variant), effectively
	suppresses MPI at all practical distances~\cite{gao2025}. The reduced
	dispersion yields 3--6$\times$ fewer FFE taps, lowering noise enhancement
	penalty and DSP power consumption.

	Economically, HCF deployment is justified now for intra-DC and campus DCI,
	where latency value and greenfield deployment economics are favorable. Metro
	DCI follows in 2027--2030 as coherent avoidance savings offset the declining
	HCF premium. The ecosystem is maturing: six or more manufacturers are scaling
	production, dedicated splicers achieve 0.043~dB mean splice
	loss~\cite{feng_ofc2026}, and production deployments exceed 1,200~km with
	15,000~km planned~\cite{microsoft2024}.
	Manufacturing capacity is scaling rapidly, and assuming standard optical industry learning curves, unit costs are estimated to reach \$1,000--2,000/km by 2030, paving a path toward price parity with premium specialty solid core fibers in the following decade.
	
	Several technical challenges remain for broader adoption. Commodity connectors
	with sub-0.3~dB loss are not yet available, limiting field deployment
	flexibility. Long-term reliability data for HCF cables under environmental
	stress (temperature cycling, humidity, mechanical load) are still being
	accumulated. Standardization through ITU-T and IEC is ongoing but not yet
	complete, and the lack of formal specifications can impede procurement by
	network operators with strict compliance requirements.
	
	As data rates increase toward 1.6T and 3.2T per wavelength, the dispersion
	sensitivity of IMDD systems grows with baud rate, strengthening
	the case for HCF's low-dispersion advantage. At 200~GBaud PAM4 (targeting
	1.6~Tb/s per lane), the first SMF fading null at just 10~km falls at
	$\sim$19~GHz---far below the $\sim$100~GHz half-baud bandwidth---making even
	short DCI links severely dispersion-limited on SMF. The convergence
	of AI-driven data center architectures---requiring low-latency, high-bandwidth
	GPU-to-GPU communication across campus-scale distances---further amplifies
	HCF's value proposition. HCF has transitioned from a laboratory demonstration
	platform to a production-deployed technology with a quantifiable path to cost
	parity.
	
	\section*{Disclaimer and Acknowledgment}
Forward-looking projections (e.g., adoption timelines and cost
	trajectories) are illustrative scenarios based on present information
	and may not materialize as or when stated. 
	
	The views and opinions expressed in this document belong solely to the author and do not reflect Huawei's official stance. 
	
	Generative AI was utilized to refine language and grammar.
	\balance
	\bibliographystyle{IEEEtran}
	\bibliography{ieee_network_hcf_imdd_v3}

@misc{heavy_reading2020,
  
  title        = {Operator Strategies for {5G} Transport: 2020 Heavy Reading Survey},
  howpublished = {July},
  year         = {2020},
  note         = {Available at \url{https://eu-assets.contentstack.com/v3/assets/blt23eb5bbc4124baa6/blt31b44108fd7263e1/64d5e88ba2fa60b34e1baa71/Operator-Strategies-for-5G-Transport-2020-Heavy-Reading-Survey.pdf
}, Accessed on: June 13, 2026}
}

@misc{yingfeng_2026,
  author       = {{Yingfeng Communication}},
  title        = {Fiber Optic Cable Price in 2026: Real Market Data for March},
  howpublished = {9~March},
  year         = {2025},
  note         = {Available at \url{https://yfconnectivity.com/latest-fiber-optic-cable-price/}, Accessed on: April 23, 2026}
}

@misc{splice_2026,
  author       = {{Latest Cost}},
  title        = {Fiber Optic Cable Cost Guide 2026},
  howpublished = {6~November},
  year         = {2025},
  note         = {Available at \url{https://latestcost.com/fiber-optic-cable-cost/}, Accessed on: April 23, 2026}
}

@misc{mapyourtech2026,
  author       = {{MapYourTech}},
  title        = {Is Hollow Core Fiber Ready for Practical Deployment?},
  howpublished = {Technical analysis, MapYourTech.com},
  year         = {2026},
  month        = feb,
  note         = {Available at \url{https://mapyourtech.com/is-hollow-core-fiber-ready-for-practical-deployment/}, Accessed on: April 23, 2026.
                  }
}

@misc{aws_dck_2026,
  author       = {Rehder, Matt},
  title        = {{AWS} Networking Boss on {Hollow-Core Fiber, AI, and Data Center Innovations}},
  howpublished = {Data Center Knowledge, interview},
  year         = {2026},
  note         = {Available at \url{https://www.datacenterknowledge.com/networking/aws-networking-boss-on-hollow-core-fiber-ai-and-data-center-innovations}, Accessed on: April 23, 2026}
}

@techreport{ec_gigabit_2023,
  author      = {{European Commission}},
  title       = {Proposal for a Regulation on measures to reduce the cost of deploying gigabit electronic communications networks ({Gigabit Infrastructure Act})},
  institution = {European Commission},
  year        = {2023},
  number      = {COM(2023) 94 final},
  month       = feb,
  note        = {Brussels, 23 February 2023. Adopted by European Parliament 23 April 2024}
}

@article{altuna_lpor_2024,
  author  = {Altuna, R. and Jung, Y. and Petropoulos, P. and Vazquez, C.},
  title   = {Power over fiber and analog radio over fiber simultaneous transmission over long distance in single-mode, multicore, and hollow-core fibers},
  journal = {Laser \& Photonics Reviews},
  volume  = {18},
  number  = {8},
  year    = {2024},
  note    = {Art. no.\ 2400157}
}

@article{yang2025,
  author  = {Yang, Hailin and Xiang, Meng and Cheng, Wenzhuo and Zhou, Gai and Li, Jianping and Wang, Yingying and Gao, Shoufei and Ding, Wei and Fu, Songnian and Qin, Yuwen},
  title   = {800{G} low-latency photonic data-center interconnections over 5\,km hollow-core fiber},
  journal = {IEEE Communications Magazine},
  volume  = {63},
  number  = {3},
  pages   = {122--128},
  year    = {2025},
  doi     = {10.1109/MCOM.001.2300806},
}

@article{gao2025,
  author  = {Gao, Shoufei and Chen, Hao and Sun, Yizhi and Xiong, Yifan and Yang, Zijie and Zhao, Rui and Ding, Wei and Wang, Yingying},
  title   = {Fourfold truncated double-nested antiresonant hollow-core fiber with ultralow loss and ultrahigh mode purity},
  journal = {Optica},
  volume  = {12},
  number  = {1},
  pages   = {56--61},
  year    = {2025},
  doi     = {10.1364/OPTICA.542911},
}

@misc{microsoft2024,
  author = {O. Khan},
  title  = {How hollow core fiber is accelerating {AI}},
  howpublished = {Microsoft Azure Blog},
  year   = {2024},
  note   = {Available at \url{https://www.networkworld.com/article/4049666/microsofts-hollow-core-fiber-delivers-the-lowest-signal-loss-ever.html}. Accessed on: March 18, 2026},

}

@article{nyarkoboateng2020,
  author  = {Nyarko-Boateng, Owusu and Xedagbui, Faith Edem Bright and Adekoya, Adebayo Felix and Weyori, Benjamin Asubam},
  title   = {Fiber optic deployment challenges and their management in a developing country: A tutorial and case study in {Ghana}},
  journal = {Engineering Reports},
  volume  = {2},
  number  = {2},
  pages   = {e12121},
  year    = {2020},
  doi     = {10.1002/eng2.12121},
}

@misc{lightcounting2024,
  author       = {{LightCounting}},
  title        = {Optical Transceivers Forecast},
  howpublished = {Market Research Report},
  year         = {2024},
}

@misc{oif2020,
  author       = {{Optical Internetworking Forum}},
  title        = {{400ZR} Implementation Agreement ({OIF}-400{ZR}-02.0)},
  howpublished = {OIF Specification},
  year         = {2020},
  note         = {Available at \url{https://www.oiforum.com/wp-content/uploads/OIF-400ZR-02.0.pdf}.},
}

@inproceedings{feng_ofc2026,
  author    = {L. Feng and W. He and C. Zhang and others},
  title     = {Fast, Low-Loss, and Field-Deployable Splicing of Anti-Resonant Hollow-Core Fibers},
  booktitle = {Optical Fiber Communication Conference (OFC)},
  address   = {Los Angeles, CA, USA},
  year      = {2026},
  organization={Optica Publishing Group},
  pages     = {M1J.4},
}

@inproceedings{slavik_ofc2026,
  author    = {R. Slavik and M. Komanec and F. Poletti},
  title     = {Making Hollow Core Fibers Compatible With Current Fiber Infrastructure},
  booktitle = {Optical Fiber Communication Conference (OFC)},
  address   = {Los Angeles, CA, USA},
  year      = {2026},
  organization={Optica Publishing Group},
  pages     = {M1J.5},
}
	
\end{document}